\documentclass[12pt,a4paper,twoside]{proceedings_book}
\usepackage{graphics,references,miscellaneous}
\footnotesep=\baselineskip
\date{\thanks{Astrophysical Dynamics 1999/2000,
              Alessandro B. Romeo (Ed.),
              Onsala Space Observatory,
              2000.}}

\begin{document}

\title{The Outflow-Disc Interaction in Young Stellar Objects}

\author{Michele Pestalozzi\\[.33cm]
        Onsala Space Observatory \\
    Chalmers University of Technology \\
    SE-43992 Onsala, Sweden\\
        (michele@oso.chalmers.se)}

\pagestyle{myheadings}

\markboth{The Outflow-Disc Interaction in Young Stellar Objects}
         {Michele Pestalozzi}

\maketitle

\begin{abstract}

In this work the most spectacular phenomena occurring during the formation of a star are briefly reviewed:
accretion through a rotating disc of matter and outflow through the poles of the new stellar object. Magnetic
fields have been proposed to be principally responsible for the coexistence of these opposed mechanisms of
accretion and outflowing. According to different models, the magnetic fields are either twisted or stretched
by the accretion disc, allowing the formation of polar channels through which part of the accreting matter
can escape. The high degree of coupling between the physical parameters describing the young stellar object
(e.g. circular velocity of the disc and the central object, viscosity, strength and freezing of magnetic
fields, etc.) makes detailed understanding of the interaction between accretion discs and outflows very
difficult. The models presented here provide only a partial answer to this difficult problem.

\end{abstract}

\section{Introduction}

\begin{figure}
\begin{center}
\resizebox{9cm}{9cm}{\includegraphics{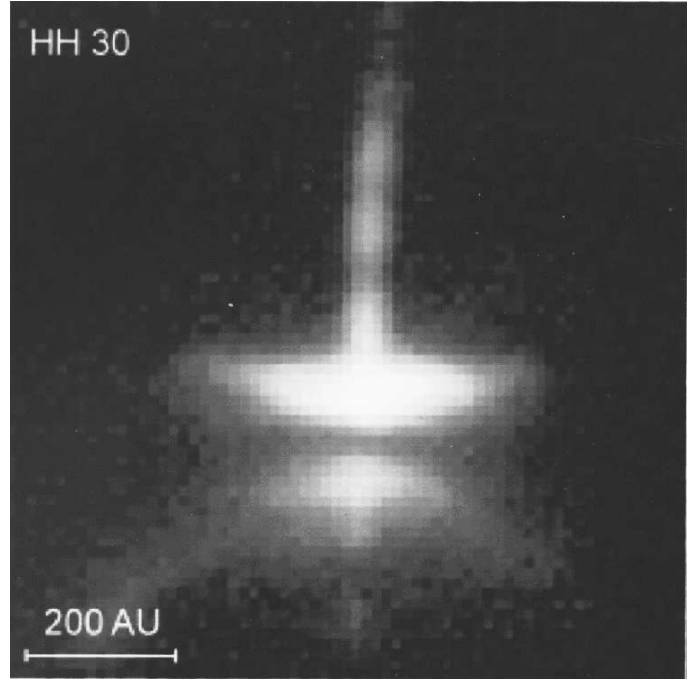}} \caption{Single wavelength Hubble Space Telescope (HST)
images do not give information about the dynamics of young stellar objects but they are able to constrain
their geometry. Here is shown HH 30, a T Tauri star, with a disc and two opposite polar outflows. Picture
taken from the NASA archive.} \label{fig:HST}
\end{center}
\end{figure}

Observing a main sequence star, like our Sun, but much further away with an optical telescope will not
impress us much, since it will appear as an unresolved point source of light, without any particular feature.
Main sequence stars are in fact objects in the most stable and quiet stage of their life. A slightly higher
degree of activity (e.g. flares) can be observed at other wavelengths (e.g. radio or X-rays): synchrotron
emission from the very hot plasma in the corona of the star is one of the most evident hints for the
existence of strong magnetic fields rooted in the star. Nevertheless this activity is not comparable to the
one through which the star had to go before reaching the quiet days of the main sequence. First the collapse
of a molecular cloud, which can be of several light years in size. Then the accretion of matter through a
disc accompanied by outflows (see Fig. \ref{fig:HST}). Finally the star reaches the stage of the ignition of
nuclear fusion in its core which leads to emission at optical wavelengths, which will characterize its life
on the main sequence.

I concentrate on the second phase mentioned above and explain, on the basis of both on theory and
observations, the coexistence of two apparently opposite phenomena as accretion and outflow of matter. To do
this I will first introduce briefly (in section 2) the physical principles which are needed for such a study.
I will then review a sample of theoretical models and the results of recent simulations (section 3). In
section 4 I will draw the best links between the model predictions and the observed phenomena. As a summary,
in section 5 I will try to delineate the further work in the field. The last section summarizes the
discussion occurred at the end of the oral presentation.

\section{Basic Equations}

{\it Magnetohydrodynamics} is the study of the motions of a fluid in the presence of magnetic fields. We
consider the material accreting and outflowing from a star as a fluid because the mean free path $l$ of a
particle is much smaller than the length of interest $L$ (in our case $l \sim 10^{14}$ cm and $L \sim
10^{19}$ cm. We therefore will need a set of equations which include both the dynamics of a fluid and the
presence of magnetic fields.

First the conservation of the {\it mass} from the hydrodynamic equations:
\begin{equation}
\frac{\partial }{\partial t} \rho + \nabla (\rho \mathbf{v}) = 0
\end{equation}

\noindent We then need an equation for the conservation of {\it momentum}:
\begin{equation}
\rho[\frac{\partial }{\partial t} \mathbf{v} +(\mathbf{v} \cdot \nabla)\mathbf{v} ] + \nabla p - \rho
\mathbf{g}= \frac{1}{4 \pi} (\nabla \times \mathbf{B})\times \mathbf{B}
\end{equation}

\noindent Notice that the term on the right marks the presence of the magnetic fields with the magnetic
force. $p$ represents the pressure and $\mathbf{g}$ the gravitational potential (which can have whatever
shape we wish).\\ \noindent The {\it energy} equation looks like the following:
\begin{equation}
\frac{\partial }{\partial t} E + \nabla \cdot (E \mathbf{v}) = -\mathbf{g} \mathbf{v} + \frac{1}{4 \pi} \eta
\mid \nabla \times \mathbf{B} \mid ^2
\end{equation}

\noindent where $\eta$ corresponds to the electric diffusivity. \\ \noindent Finally we need {\it field
equations}:
\begin{eqnarray}
\label{eq:field1} \frac{\partial }{\partial t} \mathbf{B} - \nabla \times (\mathbf{v} \times \mathbf{B}) & =
& - \nabla \times (\eta \nabla \times \mathbf{B} ) \\ \nabla \cdot \mathbf{B} & = & 0
\end{eqnarray}

This set of equations needs a few lines of explanation. First, in magnetohydrodynamics (MHD) we assume {\it
local neutrality}. Because of the very low density, the charged particles of both polarities (electrons and
ions) are supposed to fill the same space equally. This does not rule out the existence of different drift
velocities between the two species. It means that the currents exist but the electric fields can be
neglected. Hence the Lorentz force looks like the following:
\begin{equation}
\frac{1}{c} \mathbf{J} \times \mathbf{B} = \frac{1}{4 \pi} (\nabla \times \mathbf{B})\times \mathbf{B}
\end{equation}
Since the electric fields are considered small enough to be neglected, their changes in time are also
negligible. So we have
\begin{equation}
(\nabla \times \mathbf{B}) = \frac{4\pi}{c} \mathbf{J} + \frac{1}{c} \frac{\partial }{\partial t} \mathbf{E}
\end{equation}
where the last term will be equal to 0. In addition, if we rule out all dissipation effects ($\eta = 0$)
because of the very low density of the plasma, we have the right hand side of equation (\ref{eq:field1})
equal to 0. In that case we talk about field freezing, which means that the field lines follow every movement
of the matter in which they are embedded, producing changes in the geometry of the magnetic field and in the
strength of the magnetic flux.

The solution of this system of equations is a great challenge in both theoretical modelling and the numerical
computation. It will trace the magnetic field lines around young stellar objects and meet the observations
which have been carried out with all kind of techniques and at all wavelengths. In the next section we will
have a look at the models which give the best fit to the observed data.

\section{The Models}
\label{sec:models}

From the literature it is possible to recognize two major {\it schools} in modelling the magnetic fields
around young stellar objects. The first (e.g. Shu et al., 1988) considers the outflows as centrifugally
driven winds; the second stresses the twisting of the magnetic field lines.

\subsection{Magnetocentrifugally Driven Outflows}

\begin{figure}
\begin{center}
\resizebox{10cm}{7cm}{\includegraphics{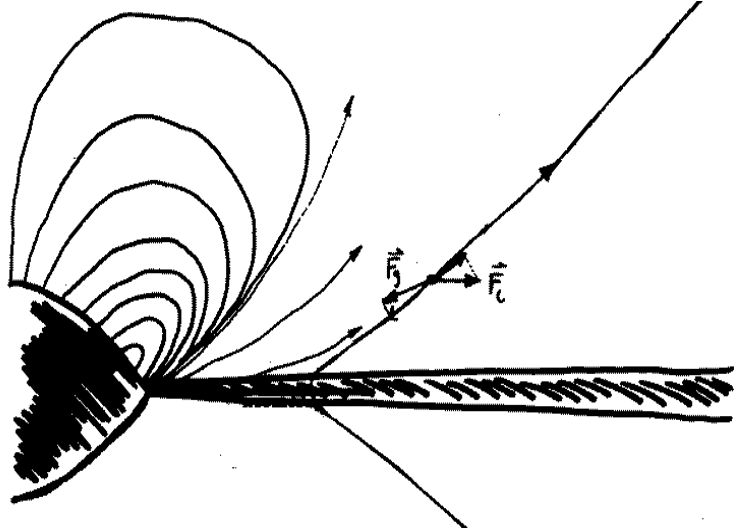}} \caption{The model of magnetocentrifugally driven outflows by
Shu et al. (1988). Note that, as a particle recedes from the star along a field line, the gravitational force
$\mathbf{F}_{g}$ becomes weaker, whereas the centrifugal force $\mathbf{F}_{c}$ becomes stronger.}
\label{fig:ShuPell}
\end{center}
\end{figure}

Shu et al. (1988, 1991, 1994) proposed a model where the outflows detectable around young stellar objects are
driven basically by the centrifugal force produced by the fast rotating central object. The latter is
supposed to rotate nearly at break up speed. The poloidal magnetic field lines are squeezed towards the
surface of the star, producing big loops expanding along polar directions. The accretion disc drives the
matter towards the star. Some of the matter travelling onto the disc surface to the star can leave the plane
of the disc and fly along the squeezed field lines. In Fig. \ref{fig:ShuPell} it is possible to see the
geometry of the field lines close to the central object and how the centrifugal force $\mathbf{F}_c$ wins
over the gravitational attraction $\mathbf{F}_g$ the further the particle travels along the field line
(Pelletier et al., 1992). The problem with this model is the collimation of the outflow. Observations show in
some cases evidence for a very efficient mechanism of collimation of the matter flying away from the central
object. This effect is not reproducible with this model. An example of an application of the model is given
in Shu et al., 1994, on SU Aur a classical T Tauri star. $M_{\star}=2.25M_{\odot}$, $R_{\star}=3.6R_{\odot}$
give a mass flow from the disc $\dot M_{D}=6\times 10^{-8} M_{\odot}/yr$ and a magnetic field strength
$B_{\star}=300 G$. The flow velocity will be around 200 km/s.

\subsection{Twisting the Magnetic Fields}

In the following models the mechanism that drives the outflows is not a wind created by the centrifugal
force. The forces acting as launching the matter away from the central object are considered to originate
only by the twisting of the magnetic field lines, which is able to open polar channels and draw matter from
the accreting disc towards the interstellar space.

\begin{figure}[!t]
\begin{center}
\resizebox{13cm}{8cm}{\includegraphics{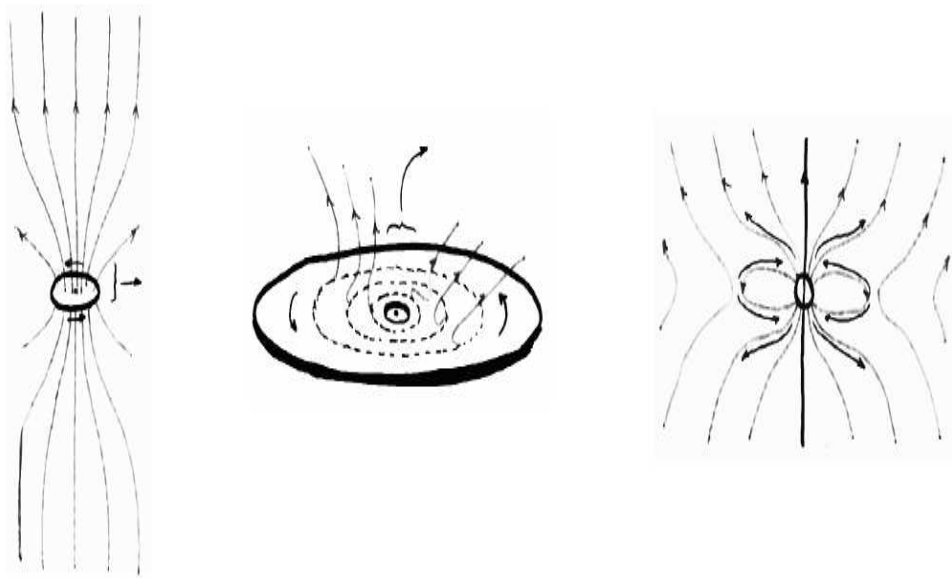}} \caption{The model by Uchida \& Shibata (1985). The three
panels show a progressive zoom into the star formation region. Note in the right panel the {\it closed} and
{\it opened} field lines.} \label{fig:Uchi}
\end{center}
\end{figure}

The first of these was proposed by Uchida \& Shibata, 1985. It considers an interstellar magnetic field,
frozen into the interstellar medium. Due to the collapse of the original molecular cloud the magnetic field
lines are constrained to come closer to the central object, and because of the rotation of the accretion disc
they will be twisted. The result is shown in Fig. \ref{fig:Uchi}. The twisting of the field leads to the
appearance of a $\mathbf{J} \times \mathbf{B}$ force: $\mathbf{J}$ is radial due to infalling material,
$\mathbf{B}$ leaves its initial configuration parallel to the rotation axis and, because of the twist due to
the rotation of the disc, lies more and more in the plane of the disc. The $\mathbf{J} \times \mathbf{B}$
force is therefore oriented perpendicular to the disc and is able to produce large scale weakly collimated
outflows. The collimated hot optical jets, on the other hand, are considered to have their origin closer to
the star: material falling to the surface of the star can be blown off in the polar direction along the
``open'' field lines (these are not strictly open, they in fact reconnect with the interstellar field).

\begin{figure}[!b]
\begin{center}
\resizebox{14cm}{6cm}{\includegraphics{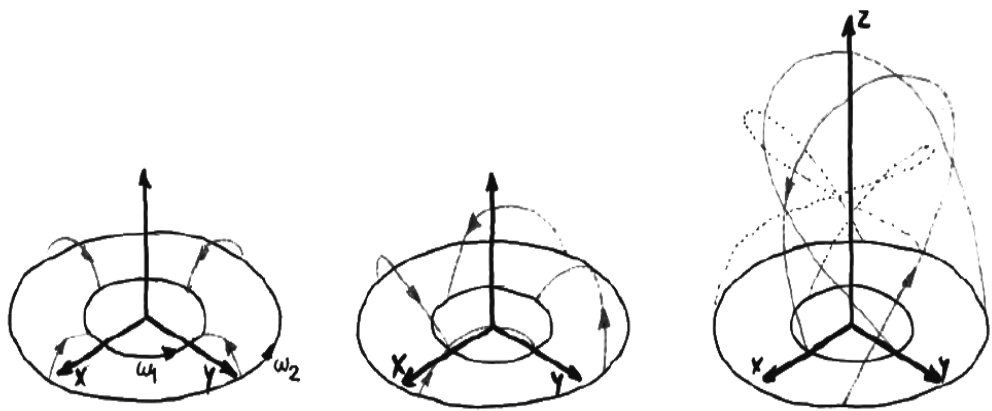}} \caption{The Newman et al. (1992) model shows how
differential rotation can twist the magnetic field lines, forcing them to expand in the polar direction. The
time sequence starts from the left drawing.} \label{fig:Lov}
\end{center}
\end{figure}

The second model comes from Newman et al. (1992) (Fig. \ref{fig:Lov}). The important assumption here is
$\mathbf{J} \times \mathbf{B} = 0$, which indicates a force free plasma. The Poynting theorem applied to this
condition shows that because of the differential rotation of the disc (footpoints of the poloidal field
travel apart form each other) the energy of the field has to increase. This energy enhancement causes the
field lines to ``inflate'', especially in the direction of the poles. Material can then escape along the
inflated field lines, away from the disc surface. Simulations by Lovelace et al. (1995) show that very little
time is needed in order to reach a reasonable field inflating, which means that the phenomenon of outflow is
present at the very beginning of the formation of the star, as soon as the accretion disc is present.

\begin{figure}
\begin{center}
\resizebox{14cm}{7cm}{\includegraphics{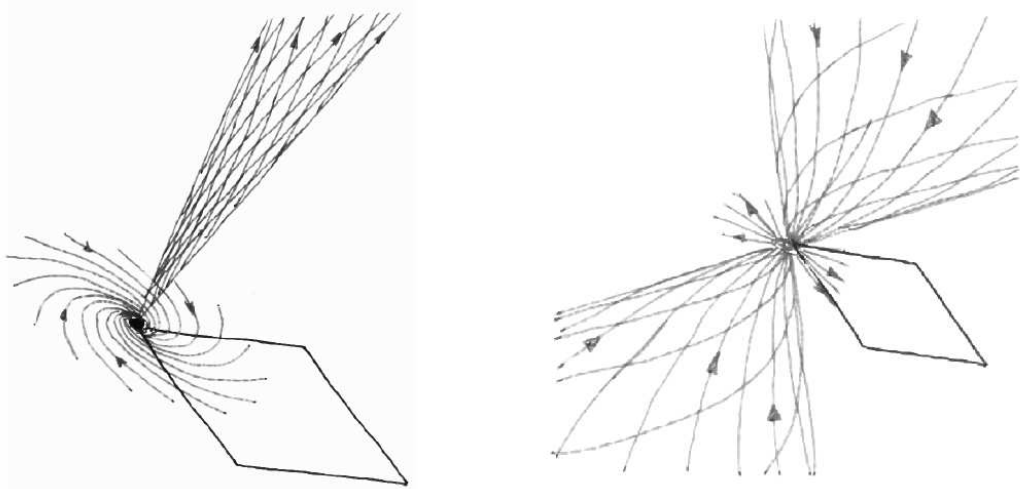}} \caption{The Henriksen \& Valls-Gabaud (1994) model lets
open a more general scenario, even a polar inflow with equatorial outflow (right panel).} \label{fig:Hen}
\end{center}
\end{figure}

The third model was proposed by Henriksen \& Valls-Gabaud (1994) and is called {\it cored apple}. It is
explicitly said in the article that this model cannot reproduce any observation with high fidelity.
Nevertheless it serves to show that no possibility can be ruled out. In particular their simulation shows,
beside a known equatorial-infall-/polar-outflow-model, the opposite scenario of having a polar inflow (better
called infall) and an outflow in the equatorial plane as (see Fig. \ref{fig:Hen}).
This mechanism is supposed to appear at the very beginning of the star formation, i.e. during the collapse of
the molecular cloud, when yet no disc-like geometry has been created from the accreting material. Only a
thick torus is present. The engine of the outflow is only the rebounding of the free falling material. It is
even possible for some of the gas to miss the central object and to be directly conducted outwards. The model
is however not able to reproduce field energies, pressures and flow velocities measured by observation,
especially when the distance to the central object becomes greater than 100 AU (tests of the model have been
carried out on VLA 1623, see Henriksen \& Valls-Gabaud, 1994).

There have also been several attempts of modelling time-dependent accretion and outflow, see e.g. Goodson et
al., 1997, 1999a, 1999b. They basically assume a radially oscillating disc, which periodically squeezes the
poloidal magnetic field (field configuration of Newman et al., 1992), which inflates and allows material to
flow out in the form of ``bullets''. The model is in good agreement with observations at different
wavelengths of HH 30 (see Fig. \ref{fig:HST}) and DG Tau, two classical T Tauri stars. Further attempts of
describing the material ejection time dependence claim to trace both the phenomena around a black hole and
around a protostar (Ouyed et al., 1997).

The latest models are able to combine the existence of a fast, hot and narrow outflow (near to the rotation
axis) with a slow outflow, colder and closer to the disc. The first would be driven by the dynamo action of
the disc which gives raise to an azimuthal field, the second would be centrifugally driven. The former would
be responsible for carrying material away from the disc, the latter to transport angular momentum outwards,
along the disc (Brandenburg et al., 2000).

\section{Observations: The Reality}

From the pictures of the HST we get a static image of a forming star (Fig. \ref{fig:HST}). All information of
the dynamics in that region is completely missing. In order to supply this information it is important to
combine different techniques. In particular, I will present some results obtained from spectrometric studies
of those regions at millimeter wavelengths. The reason for using radio observations instead of optical ones
is simple: optical radiation is absorbed by the great amount of dust present around the central object; radio
waves on the contrary get through and carry very important information about the interior of the cloud.
Infrared observations are also important, since the dust scatters the $UV$-light from the star to longer
wavelengths producing a {\it red excess}, i.e. a higher flux in the infrared than would be expected from the
star. This excess radiation gives us information about the amount of dust around the forming star.

\begin{figure}
\begin{center}
\resizebox{12cm}{7cm}{\includegraphics{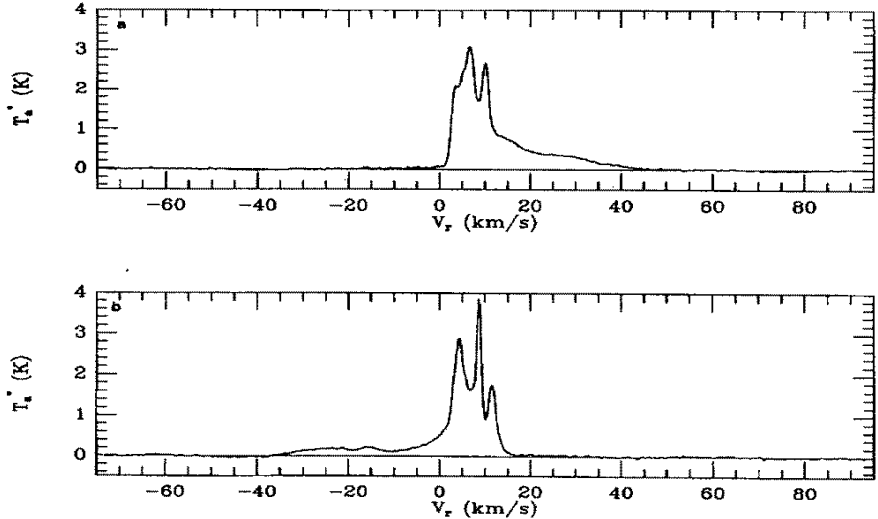}} \caption{The emission spectrum of the CO transition $J=2
\rightarrow 1$ towards the outflow in NGC 2264G. The upper panel shows the emission of the redshifted lobe of
the outflow, the lower panel the emission of the blueshifted (Lada \& Fich 1996).} \label{fig:line}
\end{center}
\end{figure}

\begin{figure}[!t]
\begin{center}
\resizebox{16cm}{11.5cm}{\includegraphics{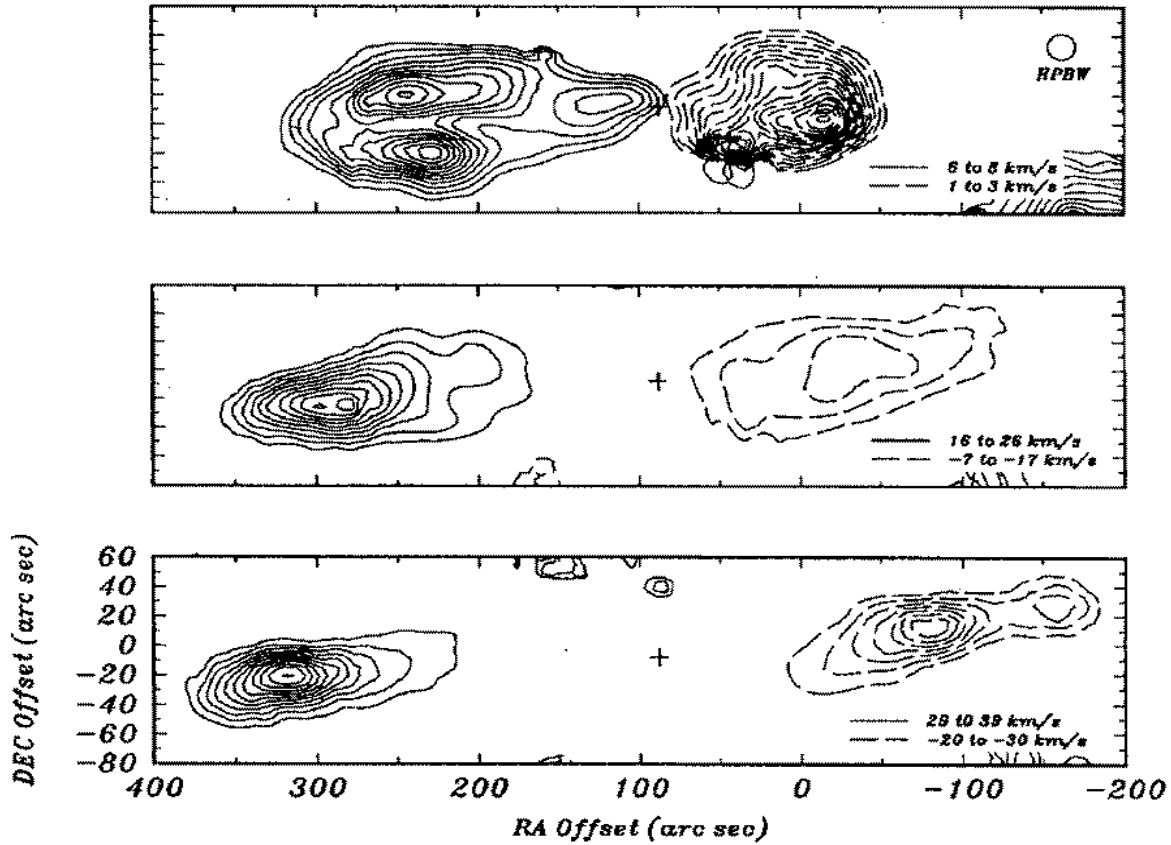}} \caption{From the spectrum above it is possible to map
the emission spatially: every slide corresponds to a range relative velocities, showing how the material is
distributed both in space and velocity (Lada \& Fich 1996).} \label{fig:outflow}
\end{center}
\end{figure}

An observed spectral line will often be shifted in frequency, because of the relative radial velocity between
us and the observed object (Doppler-effect). From the amplitude of that shift and from the shape of the
spectral line it is possible to draw conclusions about the relative motion of the matter, e.g. if it is
moving towards us or away from us. In Fig. \ref{fig:line} a typical spectrum is shown. Using an
interferometer the data can be mapped in both space and velocity to study the source dynamics. Clearly
noticeable is the velocity distribution of the matter at different distances from the central object (Fig.
\ref{fig:outflow}).

\begin{figure}
\begin{center}
\resizebox{15cm}{7cm}{\includegraphics{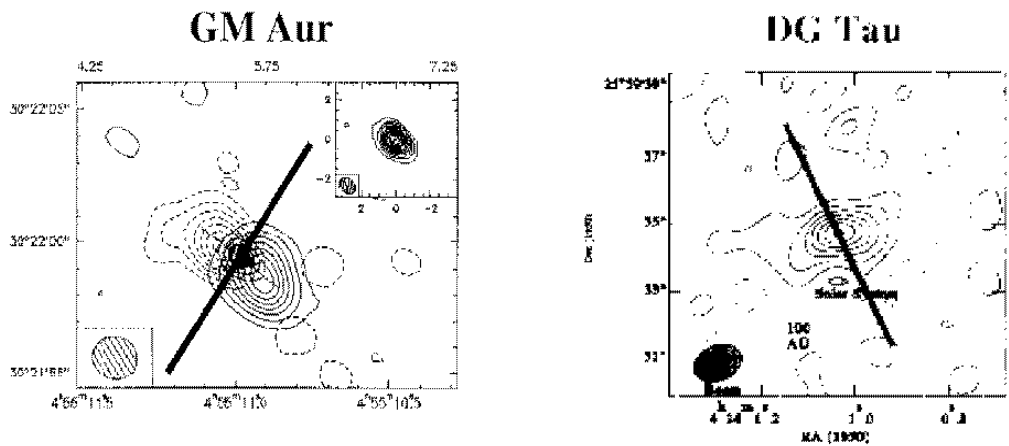}} \caption{Two examples of a measurement of the polarisation
of the mm or sub-mm continuum emission coming from young stellar objects (in this case two classical T Tauri
stars). The thick line indicates the orientation of the magnetic field, which lies perpendicular to the
outflows shown in the CO. One can conclude that the toroidal component of the magnetic field dominates the
configuration around the YSO, indicating an advanced state of evolution (the magnetic field lines have been
twisted for long time) (Tamura et al. 1999).} \label{fig:Tamura}
\end{center}
\end{figure}

But even more useful for the study of the magnetic fields is the measure of the polarization of the light. By
observing at millimeter and submillimeter wavelengths it is possible to measure the amount and direction of
polarization as e.g. shown in the paper by Tamura et al. (1999). Dust particles aligned with the magnetic
fields scatter $UV$-light from the star producing a slight polarization. In Fig. \ref{fig:Tamura} one can see
the orientation of the polarization (thick line) superimposed on CO contours which traces the outflows.
Realizing that the magnetic field is oriented along the disc, one can conclude that the toroidal component of
the magnetic field is dominating the configuration around the two objects. It is also conceivable that
determining the age of young stellar objects will be possible through measuring of the polarization: the more
alignment with the accretion disc is detected, the older is the YSO (as time goes by the configuration is
geometrically more defined and the magnetic field more twisted in the toroidal direction). A discussion about
the possible correlation between ``order'' in the magnetic field and age is presented by Greaves et al.
(1997).

\begin{figure}[!t]
\begin{center}
\resizebox{8cm}{7cm}{\includegraphics{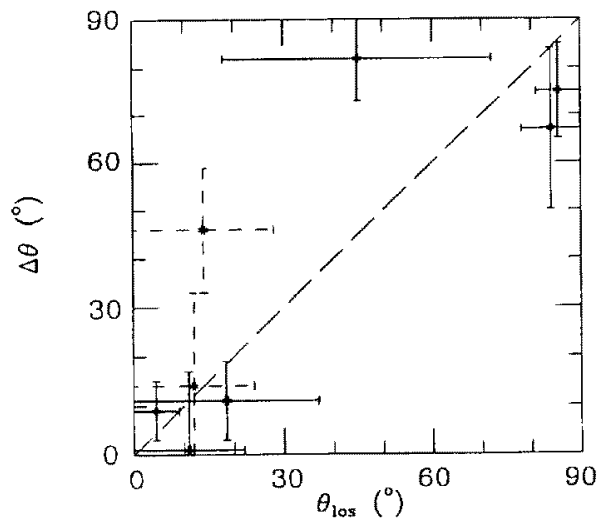}} \caption{This diagram shows the correlation between
$\Delta\Theta$ (angle between the magnetic field and the outflow direction) and $\Theta_{los}$ (angle between
the outflow and the line of sight to the observer, estimated geometrically. From this diagram it is possible
to determine the geometry of the object (Greaves et al. 1997).} \label{fig:theta_pol}
\end{center}
\end{figure}

The choice between one theoretical model and another is not always clear. Fig. \ref{fig:theta_pol} shows a
plot of polarization against viewing angle. The latter is calculated based on geometrical considerations (see
Greaves et al., 1997). The clear correlation between the polarization angle and the angle of view is striking
and this suggested to the authors different conclusions about the geometry of the star-disc-outflows system.
From the diagram in Fig. \ref{fig:theta_pol} one can see that when the outflow lies in the plane of the sky
($\Theta=90^o$) the magnetic field lies perpendicular to the outflow direction. The opposite happens when the
outflow is oriented close to the line of sight: the magnetic field seems to be parallel to the outflow. The
authors test against the last three models of section \ref{sec:models}. Though a definitive choice between
models is not possible they can nevertheless possible to rule out one of them in some cases.

\section{What Remains to Be Done}

In my opinion, the greatest challenge of the next generation of instruments (ALMA for the submillimeter,
optical and infrared interferometry, etc.) is to achieve higher spatial resolution in order to be able to
{\it zoom} into the deepest regions of the formation of a star. Polarization measurements are also well
suited to study the presence and the evolution of magnetic fields in star formation regions. All this will be
a great contribution to the modelling of both jets from black holes and outflows from YSO's. In active stars
(pre main sequence), the tracing of the magnetic fields can also be done by observing synchrotron emission
from the plasma of the corona. These studies give insight into the existence and evolution of the corona in
early phases of stellar evolution, in order to better understand the origin of e.g. the solar corona. An
ongoing project tries in fact to detect T Tauri stars at very high resolution through synchrotron emission,
with the aim of tracing the distribution of the hot plasma and consequently of the magnetic field.

\section{Questions and Discussion}

\begin{itemize}
\item {\bf Student}: Where does the outflowing matter go? Does it come back on the star later on? \\

{\bf M. Pestalozzi}: The matter is blown away from the star and will in principle not participate in the
formation of the star where it originated. Eventually it might be part of the formation of another star in
the neighbourhood. When stars form in clusters the star formation rate is triggered exactly by this kind of
phenomenon, which spreads out material into the interstellar space.

\item {\bf A. Romeo}: How does accretion occur? \\

{\bf M. Pestalozzi}: A single particle moving in a central field has a constant angular momentum. This fixes
the characteristics of the orbits that the particle can have (the shorter the radius, the higher the rotation
velocity). In order to fall towards the central object the particle has to lose angular momentum and energy,
which in a YSO occur through the frequent collisions of gas particles. Research is in progress to understand
how these dissipative phenomena are correlated.

\item {\bf A. Tappe}: Jets from a black hole and outflows from a YSO: are they the same phenomena? Will it be possible to join them in the same class? \\

{\bf M. Pestalozzi}: In the simulations by Ouyed, Pudritz \& Stone (1997) the propagation of jets and
outflows does not differ, except for the scale. However YSO and black holes are physically very different
objects, especially close to the center. The unification of the theories describing outflows or jets is
probably not possible at scales comparable to the size of the central object. Further out the morphology of
the dynamics of the outflowing material have actually similar characteristics in both type of objects.

\item {\bf M. Thomasson}: How does polarization of light occur? \\

{\bf M. Pestalozzi}: Light coming from the star scatters on the dust particles which lie around the star
(either in the disc or in general in the cocoon of the forming star). Magnetic fields on the other hand have
the capacity of orienting the dust particles along the field lines, according to their magnetic properties:
every elongated particle will feel a magnetic momentum, which will force the particle to assume the less
expensive position in terms of energy. In this situation, light which scatters with {\it oriented} dust will
orient its fields, producing a polarization along the major axis of the dust particles.

\item {\bf M. Thomasson}: Where is the force that drives the matter out of the plane (e.g. in Newman et al., 1992)? \\

{\bf M. Pestalozzi}: In the model by Uchida \& Shibata, 1985 it is the $\mathbf{J} \times \mathbf{B}$ force
which increase with the twisting of the field lines by the rotation of the disc. In Newman et al., 1992,
material follows the field lines which have been inflated by the stretching of the field lines due to
differential rotation of the disc.

\item {\bf A. Tappe}: Open field lines: is this possible? \\

{\bf M. Pestalozzi}: What is meant by {\it open field lines} are not lines which will never close, opening
the possibility to the existence of magnetic monopoles. These lines are supposed to reconnect with the lines
of the interstellar magnetic field. They are virtually open for the central object we are considering.

\item {\bf A. Tappe}: Is it possible to tell which model is right and which is wrong? And if yes, how? \\

{\bf M. Pestalozzi}: No. There are better models and worse ones. The whole thing is about how well a certain
model fits the observations. Some of the presented models are better for the explanation of some observations
but not of others (see Greaves et al., 1997).

\item {\bf V. Minier}: You are speaking about low-mass star formation. What about high-mass star formation? Is any model possible for it? \\

{\bf M. Pestalozzi}: I was actually expecting that question. In my opinion that I have been cultivating
during the preparation of this seminar, there should not be any difference between the formation of a
low-mass star and a high-mass star, except for the scale of the phenomena. The problem comes from the time
scale and the chance to observe a star in this phase: high-mass stars are supposed to form and evolve quicker
than low-mass stars. This already puts a constraint on the {\it chance} of finding a high-mass star exactly
in the evolutionary phase we would like. In addition, there are many more low-mass stars than high-mass ones.
In any case, basing our consideration on the existent literature, I suggest to look at a high mass star
formation applying Henriksen's model. Since it seems to be difficult for a 50 M$_{\odot}$ star to keep intact
discs or other nice features, the model of outflows created only by the accretion seems to me the most
appropriate to describe the early phase of evolution.

\item {\bf A. Tappe}: The collimation is proportional to the twisting, or better to the velocity difference between the rings where the footpoints of the magnetic field lie. Is that correct? \\

{\bf M. Pestalozzi}: Yes. It is also well presented in the article to which I was referring. Actually in the
proposed model the authors needed to turn the simulation only about a bit more than one turn in order to get
those very pinched field lines.

\item {\bf D. de Mello}: On the HST viewgraphs you showed at the beginning a lengthscale is indicated. Could you compare it to the one you get observing with VLBI? \\

{\bf M. Pestalozzi}: In fact with VLBI observation we are able to zoom in by about two orders of magnitude
inside the objects we see in the HST picture. This seems to do all simulations about the magnetic fields
useless and a comparison between HST images and VLBI data nonsense. In fact the simulations try to reveal
real mechanism that starts the outflows, almost at the stellar surface. What we see with the HST images is a
constraint on the geometry at a larger scale. The models must then be able e.g. to collimate the outflows
within the distance form the central object that HST images indicate. Also the accretion mechanisms must obey
those geometrical boundary conditions. In such complex problems of MHD in YSOs where all effects seem to
couple (viscosity, accretion, outflow, magnetic fields, rotation, diffusion, \dots) it is important to have
at least some kind of boundary conditions on the geometry on which you can build a model.

\end{itemize}

\end{document}